\documentclass[10pt,twocolumn,letterpaper]{article}

\usepackage{cvpr}              

\usepackage{graphicx}
\usepackage{amsmath}
\usepackage{amssymb}
\usepackage{booktabs}
\usepackage{makecell}

\usepackage{graphbox}

\usepackage[accsupp]{axessibility}

\usepackage[pagebackref,breaklinks,colorlinks]{hyperref}

\usepackage[capitalize]{cleveref}
\crefname{section}{Sec.}{Secs.}
\Crefname{section}{Section}{Sections}
\Crefname{table}{Table}{Tables}
\crefname{table}{Tab.}{Tabs.}

\def\cvprPaperID{74} 
\def\confName{CVPR}
\def\confYear{2023}

\begin{document}

\title{AsConvSR: Fast and Lightweight Super-Resolution Network \\
with Assembled Convolutions}

\author{
Jiaming Guo$^{1}$ \hspace{0.5cm}Xueyi Zou$^{1}$ \hspace{0.5cm}Yuyi Chen$^{1}$ \hspace{0.5cm}Yi Liu$^{1}$ \hspace{0.5cm}Jia Hao$^{2}$ \\
\hspace{0.5cm}Jianzhuang Liu$^{1}$ \hspace{0.5cm}Youliang Yan$^{1}$\\
$^{1}$Huawei Noah's Ark Lab\\
$^{2}$HiSilicon (Shanghai) Technologies Co., Ltd\\
{\tt\small \{guojiaming5, zouxueyi, chenyuyi1, liuyi113, hao.jia, liu.jianzhuang, yanyouliang\}@huawei.com}
}
\maketitle


\begin{abstract}
In recent years, videos and images in 720p (HD), 1080p (FHD) and 4K (UHD) resolution have become more popular for display devices such as TVs, mobile phones and VR. However, these high resolution images cannot achieve the expected visual effect due to the limitation of the internet bandwidth, and bring a great challenge for super-resolution networks to achieve real-time performance. Following this challenge, we explore multiple efficient network designs, such as pixel-unshuffle, repeat upscaling, and local skip connection removal, and propose a fast and lightweight super-resolution network. Furthermore, by analyzing the applications of the idea of divide-and-conquer in super-resolution, we propose assembled convolutions which can adapt convolution kernels according to the input features. Experiments suggest that our method outperforms all the state-of-the-art efficient super-resolution models, and achieves optimal results in terms of runtime and quality. In addition, our method also wins the first place in NTIRE 2023 Real-Time Super-Resolution - Track 1 ($\times$2). The code will be available at \href{https://gitee.com/mindspore/models/tree/master/research/cv/AsConvSR}{https://gitee.com/mindspore/models /tree/master/research/cv/AsConvSR}
\end{abstract}


\section{Introduction}
Super-resolution is widely used to improve the visual quality of images and videos\cite{li2022ntire,ignatov2023efficient,ignatov2021real} displayed on various devices like mobile phones, TVs and so on. In recent years, most media contents are produced and distributed in high resolutions like 720p (HD), 1080p (FHD) and 4k (UHD), and display devices with higher resolutions have become more affordable and popular for the public. Therefore, super-resolution needs to process high-resolution images and videos , which significantly increase the processing time and memory bandwidth\cite{zamfir2023rtsr}.

Previous works like RFDN \cite{liu2020residual} and RTSRN \cite{gankhuyag2022skip} have proposed efficient networks for super-resolution, but they mainly aim at lower input resolutions like 540p and 640p. Their real-time (30fps, below 33ms per image) performance cannot be guaranteed on higher input resolutions. To design real-time super-resolution networks for inputs with higher resolutions above 720p, the effectiveness of skip-connection, concatenation and other operations which are commonly used in existing methods needs to be re-evaluated.

\begin{figure}[t]
  \centering
    \includegraphics[width=8cm, height=5cm, angle=0, scale=1]{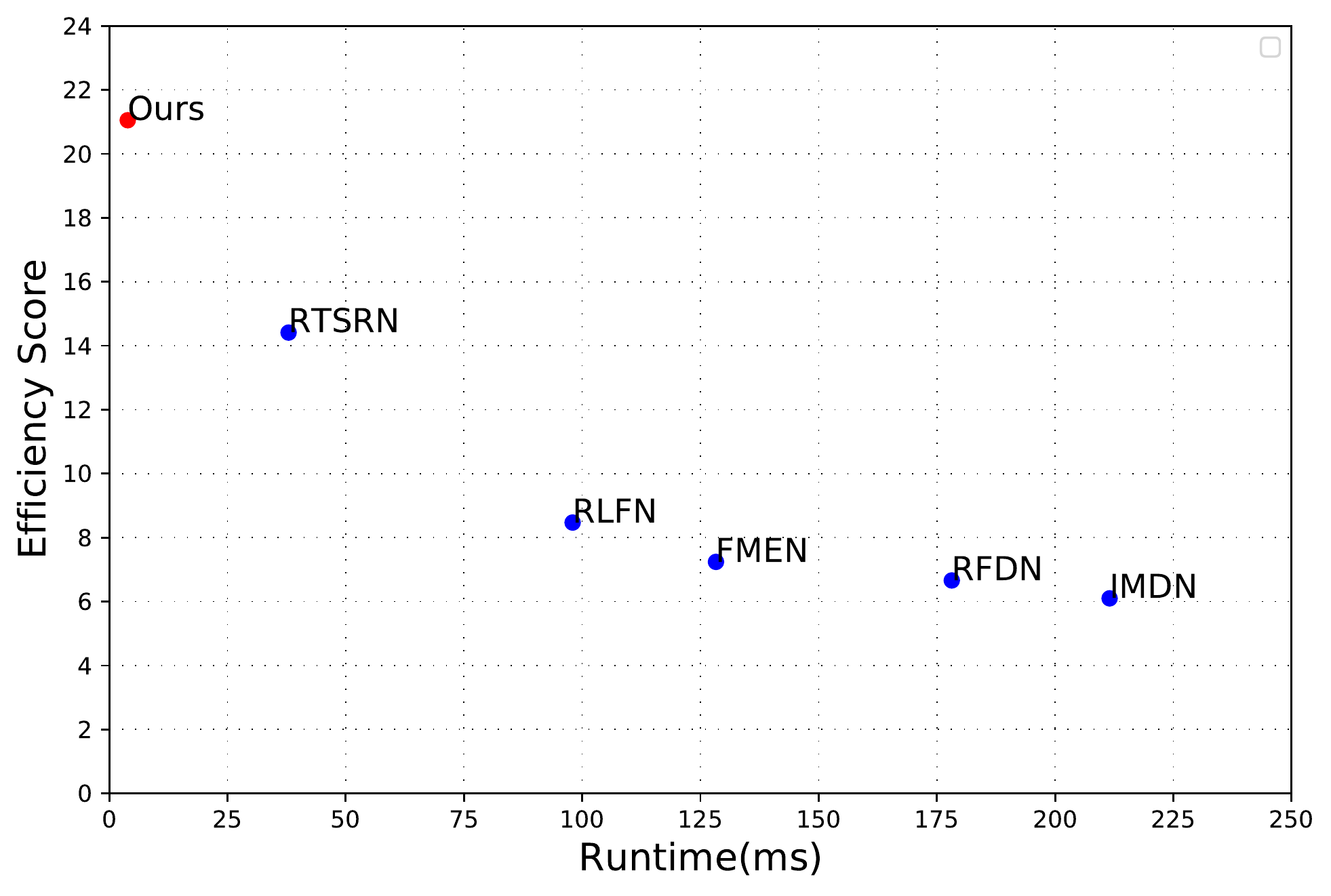}
   \caption{Illustration of efficiency score (defined in Section 4.1) and runtime of different efficient super-resolution models on DF2K dataset. Our proposed AsConvSR achieves the best performance.}
   \label{fig:1}
\end{figure}

Facing the aforementioned challenge, we redesign the basic structure of the network to achieve the goal of real-time super-resolution on high input resolutions like 720p and 1080p. We re-evaluate the performances of the network structures with complex topologies such as Enhanced Spatial Attention (ESA) and Residual Feature Distillation Block (RFDB)\cite{liu2020residual}. These structures can improve the performance of the SR network, but they also increase the model runtime. Therefore, a network with simple topology should be the best candidate to construct an efficient super-resolution model. We adopt a simple backbone that extracts features via sequential convolutions in a straight-forward topology, additionally with pixel-unshuffle and pixel-shuffle operations respectively at the beginning and the end of the network. We discard all intermediate skip-connections which bring additional computation overhead in practice, and only retain a global skip-connection to achieve high efficiency.

Considering the images as inputs for the super-resolution network we discuss above, we usually see patches or areas with different textures and contents from a great variety of objects like trees, buildings, human beings, etc. Intuitively, these patches with different patterns and texture complexity require different processing methods. Therefore, we propose the assembled convolution which enables the networks to adaptively apply different convolution kernels for different inputs. Compared with previous dynamic convolution\cite{yang2019condconv, chen2020dynamic, zhang2020dynet}, our assembled convolution is more flexible and effective as it calculates the optimal convolution kernel coefficients for each channel, and the design only brings a slight increase in computation cost and inference time for the entire network.

As demonstrated in Fig.\ref{fig:1}, our model outperforms all the state-of-the-art efficient super-resolution model in the efficiency score and runtime. With our compact design, our model is capable of real-time super-resolution on 720p and 1080p inputs with a single GPU (NVidia Tesla V100).

Our main contributions can be summarized as follows:
\begin{itemize}
\item We propose a fast and lightweight super-resolution network in a simple and straight-forward network topology, which can keep the restoration accuracy and significantly reduce the model runtime.
\item We propose assembled convolution for our efficient super-resolution network. It expands the capacity of feature extraction and meanwhile retains efficiency by adapting convolution kernels based on the input.
\item By applying the assembled convolution in the fast and lightweight super-resolution network, we propose AsConvSR which wins the first place in NTIRE 2023 Real-Time Super-Resolution - Track 1 (X2)\cite{conde2023ntire_rtsr}.
\end{itemize}

\section{Related Work}

\subsection{Efficient Image Super-Resolution}
In recent years, many approaches have been proposed to devise efficient super-resolution networks\cite{hui2019lightweight,liu2020residual,kong2022residual,du2022fast,wang2022edge,ahn2018fast,gao2022feature} that can run robustly on devices with limited computational resources with low latency. Rethinking the pioneering work SRCNN \cite{dong2015image} that applies deep learning to SISR for the first time, FSRCNN \cite{dong2016accelerating} significantly accelerates the SISR network by adopting the original low-resolution as input without bicubic interpolation, smaller sizes of convolution kernels, and a deconvolution layer at the final stage of the network to perform upsampling. LapSRN \cite{lai2017deep} progressively reconstructs the sub-band residuals of high-resolution images using the Laplacian pyramid. CARN \cite{ahn2018fast} further improves efficiency by its design of cascading residual networks with group convolution. IMDN\cite{hui2019lightweight} proposes information multi-distillation blocks with contrast-aware attention (CCA) layer based on the information distillation mechanism, while RFDN\cite{liu2020residual} refines the architecture of RFDN with feature distillation mechanism by proposing the residual feature distillation block. Following IMDN and RFDN, both RLFN and FMEN rethink the effectiveness of applying distillation and attention mechanism in this field, and consequently adopt fully sequential CNN network architecture. Specifically, RLFN \cite{kong2022residual} redesigns RFDB by adding more channels to compensate for discarded feature distillation branches to achieve higher inference speed and better performance with fewer parameters. FMEN \cite{du2022fast} expands optimization space during training with re-parameterizable building blocks \cite{ding2021repvgg} without increasing extra inference time. SwinIR\cite{liang2021swinir} proposes an efficient transformer-based SR model which fully explores the swin transformer structure, and it outperforms pure convolution networks with fewer parameters and FLOPs. Swin2SR\cite{conde2023swin2sr} further improves the network structure by introducing the SwinV2 attention, and proposes auxiliary loss and high-frequency loss for the compressed images. However, all these methods cannot satisfy real-time performance.


\begin{figure*}[t]
\centering
\includegraphics[width=18.5cm, height=6cm, angle=0, scale=1]{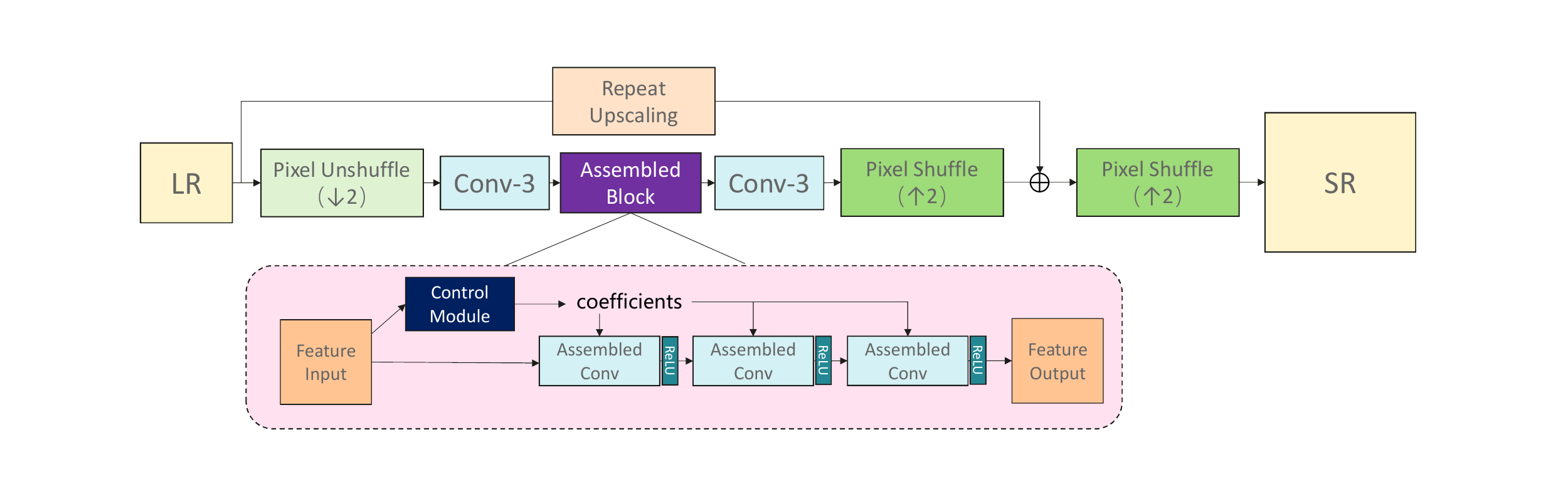}
\caption{Architecture of fast and lightweight super-resolution network with assembled convolutions (AsConvSR).}
\label{fig:2}
\end{figure*}

\subsection{Dynamic Convolution}
Dynamic Convolution\cite{bolukbasi2017adaptive,mcgill2017deciding,dehghani2018universal,wang2018skipnet,jo2018deep,xu2018dynamic,diba2019dynamonet,wu2020dynamic,zheng2020dynamic,wang2019elastic,yang2019dynamic,chen2021dynamic,wu2018dynamic} has gained growing interests for its potential to learn more powerful features by generating convolution kernels adaptively based on the inputs. Consequently, this method successfully increases capacity and flexibility of convolutional neural networks (CNN) without adding redundant computation cost, and can be readily applied in various CNN architectures. This method usually aggregates several fixed convolution kernels as introduced in \cite{yang2019condconv, chen2020dynamic, zhang2020dynet}, which also carry out experiments to demonstrate its effectiveness on typical computer vision tasks including image classification, object detection and segmentation. Dynamic convolution is also powerful on low-level vision tasks, such as denoising \cite{mildenhall2018burst}, and single image super-resolution (SISR) \cite{xu2020unified}. Inspired by the above research of dynamic convolution, we design the assembled convolution block for our efficient super-resolution network. It differs from the previous dynamic convolution, and gains higher performance as well as efficiency in super-resolution task.

\section{Method}
In this section, we first describe the network architecture of AsConvSR in Section 3.1. In Section 3.2, we analyze the dynamic convolutions and propose our assembled convolutions.

\subsection{Network Architecture}
Processing time and memory bandwidth of a neural network are highly related to the resolution of input images. RTSRN\cite{gankhuyag2022skip} can achieve real-time performance for a 360p input on a Tesla V100 GPU. However, runtime of this model would increase to 37.91 ms (see Section 4.2) when the input resolution is 1080p. The definition of floating point operations per second (FLOPs) is described as follows:

\begin{equation}
\begin{split}
    FLOPs=(\overbrace{C_{i}\times K^{2}}^{multiplications} + \overbrace{C_{i}\times K^{2}-1}^{additions}) \\
\times H\times W\times C_{o},
    \label{eq:1}
\end{split}
\end{equation}
where $C_{i}$ indicates the number of input channels, $C_{o}$ is the number of output channels, $H$ and $W$ represent the height and width of the input feature map, and $K$ is the size of the convolution kernel. Note that this formula represents the FLOPs for a convolution layer without bias. According to this equation, given a fixed FLOPs, larger $H$ and $W$ mean smaller $C_{i}$ and $C_{o}$. However, in practice, the runtime cannot be reduced by cutting down the channel size when the number of channels is smaller than a certain level (16 or 32) for most computing devices. Small channel size also limits the flexibility of design and the performance of the SR network.

\begin{figure}[h]
\centering
\includegraphics[width=8.5cm, height=3cm, angle=0, scale=1]{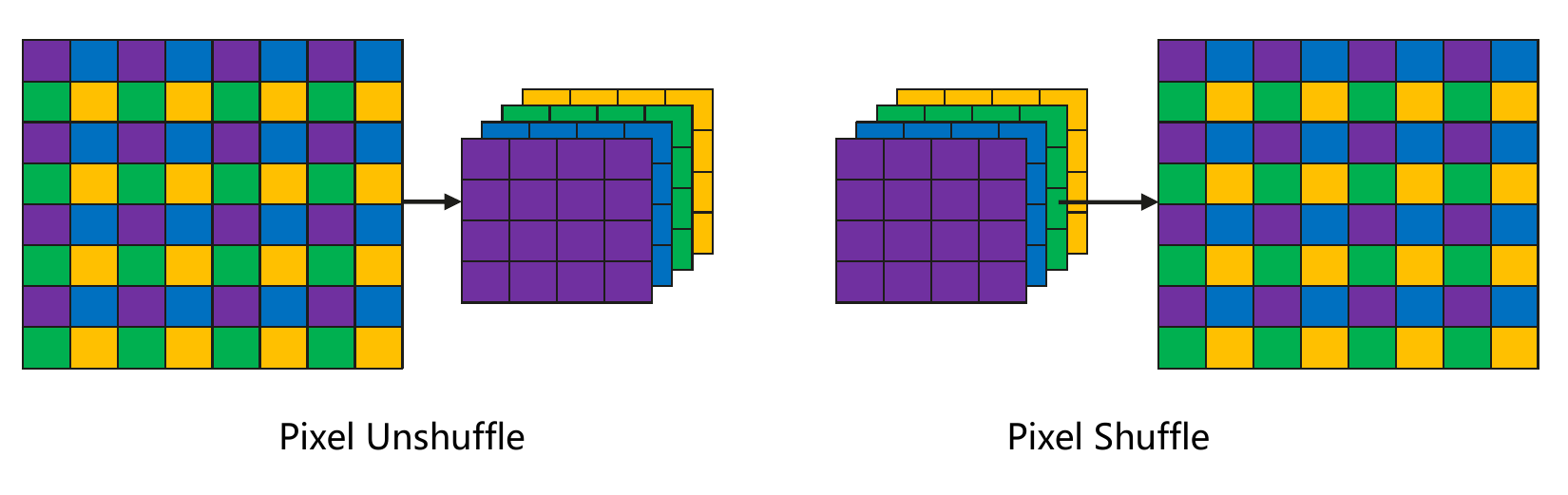}
\caption{Illustrations of pixel-unshuffle and pixel-shuffle.}
\label{fig:3}
\end{figure}

\textbf{Pixel unshuffle}. We use pixel-unshuffle operation to downscale the input image and increase the number of channels. It can reduce the computational cost of the network without loss of information volume. As demonstrated in Fig.\ref{fig:3}, pixel-unshuffle is the reverse of pixel shuffle. It rearranges the elements from input and outputs the data with smaller resolution and larger channel size. Take $N$ times pixel-unshuffle as an example, the resolution of input image downscales to $\left ( \frac{H}{N} ,\frac{W}{N}  \right )$ while the channel size becomes $C\times N^{2}$. According to Eq.\ref{eq:1}, adopting pixel-unshuffle does not change the FLOPs of the first convolution layer if the output channel size remains unchanged. From the second convolution, FLOPs decreases by a factor of $ N^{2}$ since the resolution is downscaled. The performance usually degrades if we simply downscale the input by only applying pixel-unshuffle, but this can be compensated by increasing the complexity of our network design by several options like increasing the channel size and so on.

\textbf{Skip connection}. Besides the convolution layers, skip connections also occupy a large amount of computational resources. They not only require additions of two feature maps, but also have to cache the previous feature map which increases the runtime of accessing the memory. Skip connections used in super-resolution are mainly divided into two types: one is the skip connection inside the network structure, such as the skip connection of the residual block, and the other is the global skip connection which adds up the network output and the LR upscaled by classical interpolation like bicubic or bilinear. Skip connections in the network structure can be removed according to the experiments (Section 4.4). We cannot simply remove the global skip connection because it can stabilize the training of the network and accelerate the convergence (Section 4.4). So we replace the classical interpolation algorithm in the global skip connection by repeating the LR 4 times according to ABPN\cite{du2021anchor} which is much faster than the classical interpolation.


Based on the preceding analysis, we design a fast and lightweight super-resolution network. Given an input LR image, the resolution is converted to channel dimension by pixel-unshuffle layer. By using a 3x3 convolution, the channel of feature map is converted to the target size and then fed into the assembled block. The assembled block can adaptively apply different convolution kernels for different inputs. The details about the assembled block are described in the next section. After the assembled block, a 3$\times$3 convolution layer is used to convert the channel size to 48 so that the feature map can be restored to target resolution after the pixel-shuffle layer. Noting that a low resolution image repeated in the channel dimension can also be restored to the high resolution with a pixel-shuffle layer, we divide the final pixel shuffle into two steps in order to introduce the global skip connection to the network.

\subsection{Assembled Block}
The idea of divide-and-conquer is widely used in image processing domain from classical methods to deep learning algorithms. For example, remove the noise and compression effect on flat areas, sharpen the edges in edge-dominant areas,  and generate more fine details in rich-textured areas. These intuitions underlie the idea of patched-based super-resolution. Kong et al.\cite{kong2021classsr} propose ClassSR with a classification network to determine whether the patches go to the simple sub-network to save FLOPs or the complex ones to get a better performance. However, according to our experiments, the process of splitting  the images and recombining them to patches also significantly increases the total runtime of the network. Take a 1080p image as input, the process to assemble the patches costs 7ms in total, while the entire runtime budget is only 30ms. On the other hand, dynamic convolution causes only a small increase in runtime because the dynamic overhead is at the weight level. The major computational cost of the dynamic convolution is still the 3x3 convolution, making it perfect for super-resolution task.

\begin{figure}[t]
\centering
\includegraphics[width=8.8cm, height=7.5cm, angle=0, scale=1]{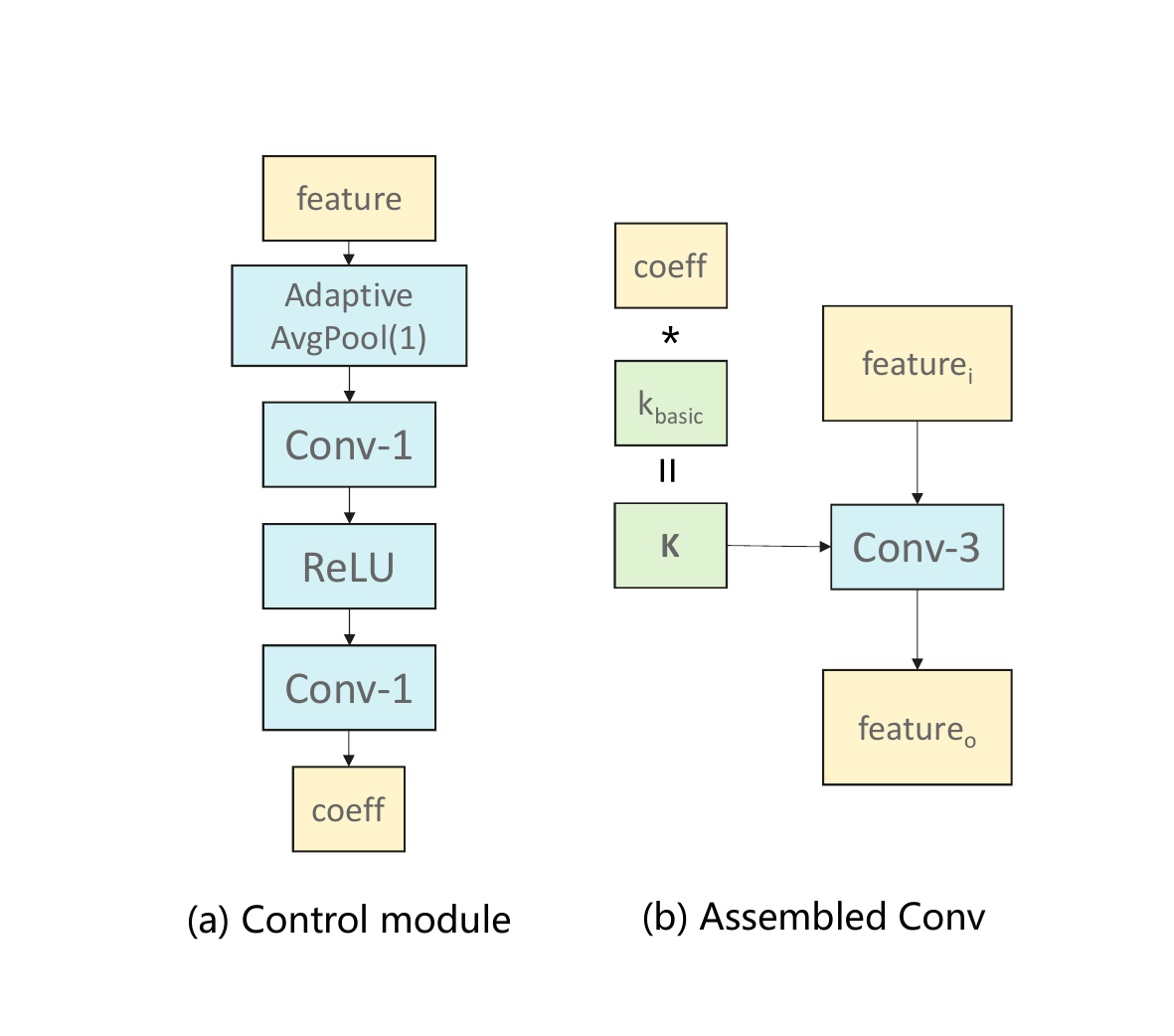}
\caption{Details of the assembled block. Conv-1 indicates a convolution whose kernel size is 1$\times$1.}
\label{fig:4}
\end{figure}

The assembled block contains a control module and three assembled convolutions. As shown in Fig.\ref{fig:4}, the control module can generate the assembled coefficients from the input feature map:
\begin{equation}
    \mathit{coeff}=F_{control}(f_{in}),
    \label{eq:2}
\end{equation}
where $f_{in}$ is the input feature, $f_{in}\in R^{B\times C_{i}\times H\times W} $, $\mathit{coeff}$ is the output coefficient, $\mathit{coeff}\in R^{B\times C_{o}\times E}$, $B$ represent the batch size, $C_{i}$ denotes the number of input channels, $C_{o}$ is the number of output channels, and $E$ is the number of candidate convolution kernels. The control module converts features into coefficients through pooling and convolution. Let
\begin{equation}
    K=\mathit{coeff}\otimes k_{basis},
    \label{eq:3}
\end{equation}
\begin{equation}
    f_{out}=conv(f_{in};K),
    \label{eq:4}
\end{equation}
where $k_{basis}\in R^{E\times C_{i}\times ks\times ks}$ is the candidate convolution kernels, $K$ is the assembled kernels for the convolution and $ks$ is the kernel size. The coefficient $\mathit{coeff}$ and candidate kernels $k_{basis}$ are first reshaped to $R^{(B \cdot  C_{o})\times E}$ and $R^{E \times (C_{i} \cdot  ks \cdot  ks)}$ respectively. Then, matrix multiplication is performed over the coefficient $\mathit{coeff}$ and candidate convolution kernels $k_{basis}$, to generate a final convolution kernel $K\in R^{B\times C_{o}\times C_{i}\times ks\times ks}$. Different batches of data require different convolution kernels, the batch dimension of the feature map is reshaped to the channel dimension and the group convolution is used to calculate the output feature maps.

\begin{figure}[h]
\centering
\includegraphics[width=8.5cm, height=6.3cm, angle=0, scale=1]{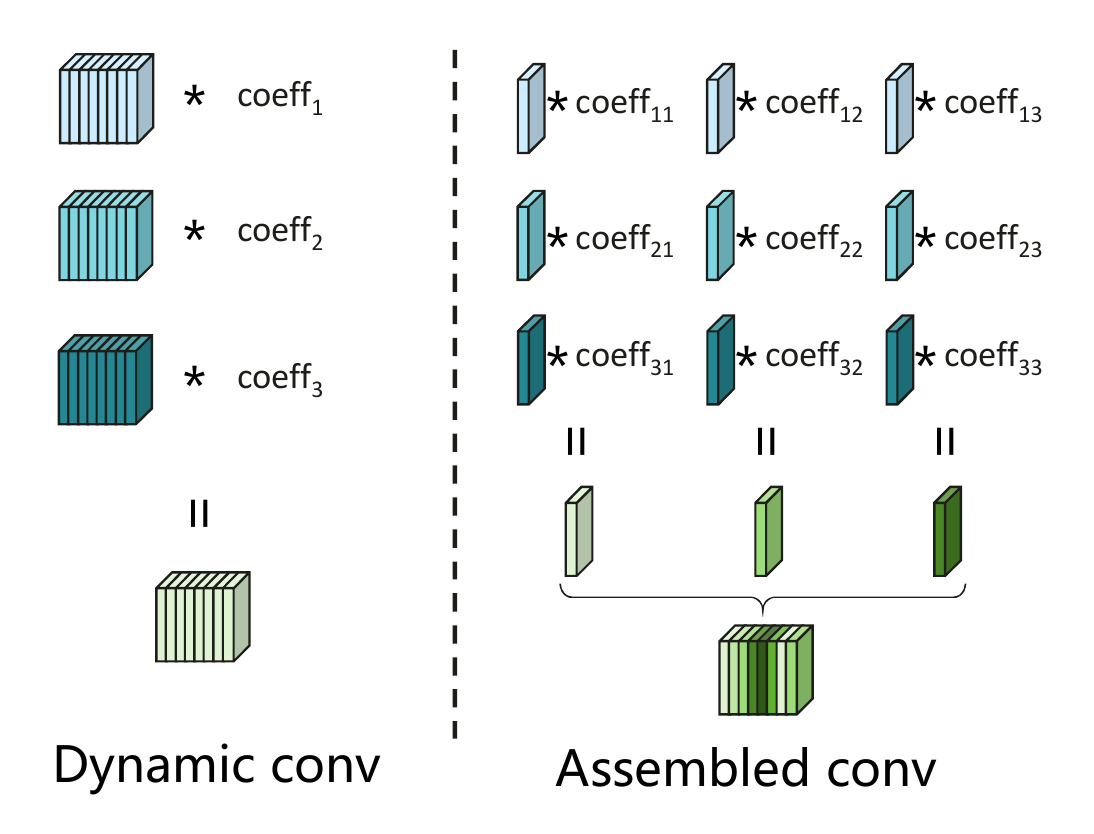}
\caption{Comparison between dynamic convolution and assembled convolution.}
\label{fig:5}
\end{figure}

Comparison between dynamic convolution and assembled convolution is shown in Fig.\ref{fig:5}. Dynamic convolution generates the whole convolution kernel (all channels) in a linear combination of the bases, which can be expressed as:

\begin{equation}
K_{dynamic}=\sum_{i=0}^{E} \mathit{coeff}_{i}\times  k_{i}^{dy},\ k_{i}^{dy}\in R^{C_{o}\times C_{i}\times ks\times ks},
\label{eq:5}
\end{equation}
where $K_{dynamic}$ is the kernel for convolution, $K_{dynamic}\in R^{C_{o}\times C_{c}\times ks\times ks}$, and $k_{i}^{dy}$ is the candidate kernel which has the same dimension as $K_{dynamic}$. Assembled convolution generates an optimal convolution kernel coefficient for each channel, which can be expressed as:

\begin{equation}
K_{j}^{rs}=\sum_{i=0}^{E} \mathit{coeff}_{i,j}\times k_{i}^{rs},\ k_{i}^{rs}\in R^{C_{i}\times ks\times ks},
\label{eq:6}
\end{equation}
\begin{equation}
K_{assembled}=cat\left ( K_{0}^{rs}, K_{1}^{rs},...,K_{C_{o} }^{rs} \right ) ,
\label{eq:7}
\end{equation}
where $K_{j}^{rs}$ is the kernel for the output channel $j$, $ K_{j}^{rs}\in R^{C_{c},ks,ks}$, $k_{i}^{dy}$ is the basis of the assembled kernel. $K_{assembled}$ for the convolution is assembled by concatenating all the $K_{j}^{rs}$. Compared with dynamic convolution, the assembled convolution we propose has finer granularity and higher flexibility in parameter generation, leading to a better performance in super-resolution.

\section{Experiments}

\subsection{Setup}
We use 3450 images in DIV2K\cite{agustsson2017ntire} and Flick2K\cite{lim2017enhanced} (DF2K) datasets for training, and test the performance of our model on five benchmark datasets: DF2K (100 testing images), Set5\cite{bevilacqua2012low}, Set14\cite{zeyde2012single}, BSD100\cite{martin2001database} and Urban100\cite{huang2015single}. We evaluate PSNR and SSIM on RGB color space. Following the ranking criterion of NTIRE 2023 Real-Time Super-Resolution challenge, we measure the efficiency of a model by the following score function:
\begin{equation}
    Score=\frac{2^{PSNR-bicubic}\cdot 2}{C\cdot \sqrt{runtime} },
    \label{eq:8}
\end{equation}
where 'bicubic' means the PSNR of the bicubic interpolation, $C$ is a constant set to 0.1 in our experiments. We adopt Adam optimizer with $\beta_{1}=0.9$ and $\beta_{2}=0.9999$ to train our model. The learning rate is $5\times10^{-4}$ in the initial stage and is halved for every $2\times10^{5}$ iterations. The entire training process takes $1\times10^{6}$ iterations to minimize the charbonnier loss. We randomly crop HR and LR patchs of sizes $256\times256$ and $128\times128$ from the training set. The mini-batch size is 32. We augment the dataset by rotating and flipping the patches of the training pairs. We use MindSpore\cite{MindSpore} and PyTorch\cite{paszke2019pytorch} for the implementation

\subsection{Quantitative Results}
\begin{table}[h]
\begin{center}
\begin{tabular}{ccccc}
    \hline
    Method & FLOPs(G) & Params(M)  & FP16 \\
    \hline
    IMDN\cite{hui2019lightweight} & 1808.98 & 0.87 & 211.52 \\
    RFDN\cite{liu2020residual}    & 824.04  & 0.42 & 178.15 \\
    RLFN\cite{kong2022residual}   & 592.36  & 0.30 & 97.98 \\
    FMEN\cite{du2022fast}         & 671.36  & 0.32 & 128.29 \\
    RTSRN\cite{gankhuyag2022skip} & 400.65  & 0.19 & 37.91 \\
    \hline
    AsConvSR-L & 36.67 & 5.21 & 24.61 \\
    AsConvSR & 9.06 & 2.35 & 3.91 \\
    \hline
\end{tabular}
\caption{FP16 indicates the runtime of a model running on half-precision floating-point in milliseconds (ms). And these runtimes are measured on a V100 GPU with 1920$\times$1080 images as inputs. We also record the FLOPs and the number of parameters of each model}
\label{tab:1}
\vspace{-0.5cm}
\end{center}
\end{table}

\begin{table*}[t]\small
\begin{center}
\begin{tabular}{ccccccc}
    \hline
    Method &
    \makecell[c]{DF2K\\PSNR/SSIM/score} &
    \makecell[c]{Set5\\PSNR/SSIM/score} &
    \makecell[c]{Set14\\PSNR/SSIM/score} &
    \makecell[c]{BSD100\\PSNR/SSIM/score} &
    \makecell[c]{Urban100\\PSNR/SSIM/score} \\
    \hline
    bicubic                       & 29.81/0.8573         & 29.96/0.8576         & 27.38/0.7987      & 27.67/0.8022      & 24.98/0.7965 \\
    IMDN\cite{hui2019lightweight} & 31.96/0.8933/{6.10}  & 32.21/0.8934/{6.52}  & 29.37/0.8409/5.43 & 29.27/0.8404/4.16 & 28.30/0.8730/13.71 \\
    RFDN\cite{liu2020residual}    & 31.96/0.8932/{6.66}  & 32.27/0.8941/{7.39}  & 29.32/0.8404/5.73 & 29.27/0.8397/4.54 & 28.36/0.8735/15.57 \\
    RLFN\cite{kong2022residual}   & 31.88/0.8915/{8.47}  & 32.14/0.8923/{9.10}  & 29.19/0.8376/7.05 & 29.20/0.8374/5.84 & 28.16/0.8696/18.21 \\
    FMEN\cite{du2022fast}         & 31.84/0.8915/{7.24}  & 32.20/0.8935/{8.34}  & 29.22/0.8387/6.32 & 29.20/0.8389/5.11 & 28.07/0.8694/15.05 \\
    RTSRN\cite{gankhuyag2022skip} & 31.52/0.8871/{10.67} & 31.93/0.8886/{12.67} & 28.96/0.8335/9.68 & 29.01/0.8345/8.20 & 27.49/0.8581/18.50 \\
    \hline
    AsConvSR-L & 31.62/0.8872/{\color{blue}14.12} & 31.95/0.8888/{\color{blue}15.98} & 29.01/0.8337/{\color{blue}12.43} & 29.05/0.8351/{\color{blue}10.46} & 27.65/0.8616/{\color{blue}25.68} \\
    AsConvSR   & 30.87/0.8766/{\color{red}21.05}  & 31.33/0.8797/{\color{red}26.00}  & 28.37/0.8202/{\color{red}19.96}  & 28.60/0.8249/{\color{red}19.23}  & 26.50/0.8365/{\color{red}28.98} \\
   \hline
\end{tabular}
\caption{Quantitative results on DF2K\cite{agustsson2017ntire,lim2017enhanced}, Set5\cite{bevilacqua2012low}, Set14\cite{zeyde2012single}, BSD100\cite{martin2001database} and Urban100\cite{huang2015single} $(\times2)$. {\color{red} Red} indicates the best and {\color{blue} blue} indicates the second best. The scores in this table are calculated by using the Eq.\ref{eq:8}. And the runtime of each model is obtained from Tab.\ref{tab:1}}
\label{tab:2}
\vspace{-0.5cm}
\end{center}
\end{table*}

\begin{figure*}[t]
\centering
    \begin{subfigure}[align=c,b]{0.33\textwidth}
        \centering\includegraphics[width=\textwidth]{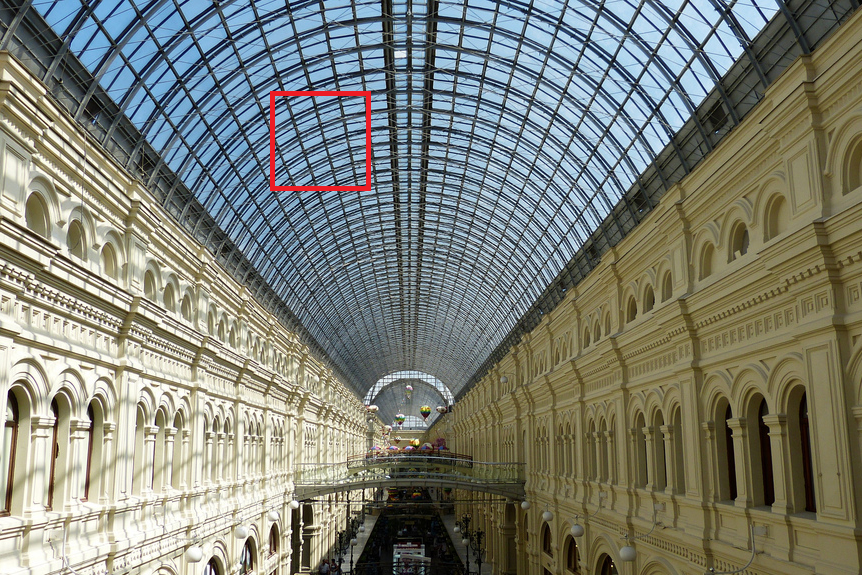}
    \end{subfigure}
    \begin{subfigure}[align=c,b]{0.65\textwidth}
        \begin{subfigure}[b]{0.23\textwidth}\includegraphics[width=\textwidth]{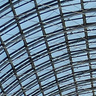}\caption*{HR}\end{subfigure}
        \begin{subfigure}[b]{0.23\textwidth}\includegraphics[width=\textwidth]{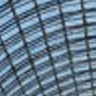}\caption*{Bicubic}\end{subfigure}
        \begin{subfigure}[b]{0.23\textwidth}\includegraphics[width=\textwidth]{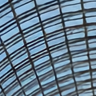}\caption*{IMDN\cite{hui2019lightweight}}\end{subfigure}
        \begin{subfigure}[b]{0.23\textwidth}\includegraphics[width=\textwidth]{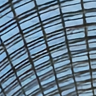}\caption*{RFDN\cite{liu2020residual}}\end{subfigure}

        \begin{subfigure}[b]{0.23\textwidth}\includegraphics[width=\textwidth]{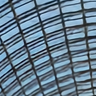}\caption*{RLFN\cite{kong2022residual}}\end{subfigure}
        \begin{subfigure}[b]{0.23\textwidth}\includegraphics[width=\textwidth]{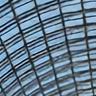}\caption*{FMEN\cite{du2022fast}}\end{subfigure}
        \begin{subfigure}[b]{0.23\textwidth}\includegraphics[width=\textwidth]{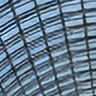}\caption*{RTSRN\cite{gankhuyag2022skip}}\end{subfigure}
        \begin{subfigure}[b]{0.23\textwidth}\includegraphics[width=\textwidth]{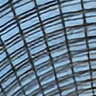}\caption*{AsConvSR-L}\end{subfigure}
    \end{subfigure}

    \begin{subfigure}[align=c,b]{0.33\textwidth}
        \centering\includegraphics[width=\textwidth]{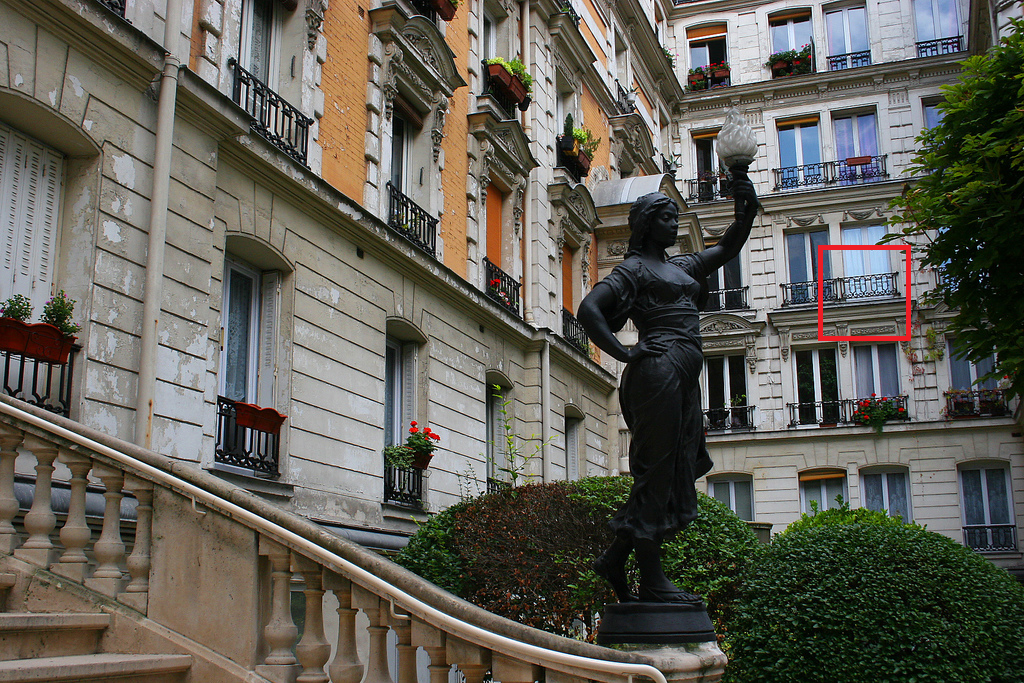}
    \end{subfigure}
    \begin{subfigure}[align=c,b]{0.65\textwidth}
        \begin{subfigure}[b]{0.23\textwidth}\includegraphics[width=\textwidth]{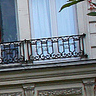}\caption*{HR}\end{subfigure}
        \begin{subfigure}[b]{0.23\textwidth}\includegraphics[width=\textwidth]{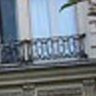}\caption*{Bicubic}\end{subfigure}
        \begin{subfigure}[b]{0.23\textwidth}\includegraphics[width=\textwidth]{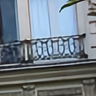}\caption*{IMDN\cite{hui2019lightweight}}\end{subfigure}
        \begin{subfigure}[b]{0.23\textwidth}\includegraphics[width=\textwidth]{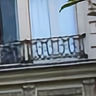}\caption*{RFDN\cite{liu2020residual}}\end{subfigure}

        \begin{subfigure}[b]{0.23\textwidth}\includegraphics[width=\textwidth]{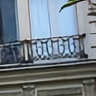}\caption*{RLFN\cite{kong2022residual}}\end{subfigure}
        \begin{subfigure}[b]{0.23\textwidth}\includegraphics[width=\textwidth]{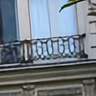}\caption*{FMEN\cite{du2022fast}}\end{subfigure}
        \begin{subfigure}[b]{0.23\textwidth}\includegraphics[width=\textwidth]{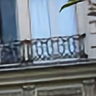}\caption*{RTSRN\cite{gankhuyag2022skip}}\end{subfigure}
        \begin{subfigure}[b]{0.23\textwidth}\includegraphics[width=\textwidth]{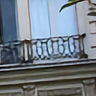}\caption*{AsConvSR-L}\end{subfigure}
    \end{subfigure}
\caption{Visual comparison of efficient SR models. Best viewed by zooming.}
\label{fig:6}
\end{figure*}

In this section, we compare AsConvSR with the state-of-the-art efficient super-resolution models. Specifically, IMDN\cite{hui2019lightweight}, RFDN\cite{liu2020residual}, RLFN\cite{kong2022residual}, FMAN\cite{du2022fast}, and RTSRN\cite{gankhuyag2022skip} are chosen for experiments. We train two versions of our model, AsConvSR-L denoting the model with a larger scale and AsConvSR a smaller one. Specifically, AsConvSR-L has two assembled blocks, each of which has 128 channels and 128 candidate kernels. And AsConvSR has only one assembled block with 32 channels.

As shown in Tab.\ref{tab:1}, none of the methods can achieve the real-time performance with 1080p input except ours. AsConvSR has only $2.26\%$ of RTSRN's FLOPs and $10.33\%$ of RTSRN's runtime. Because assembled convolution has a large number of bases of weights, the number of parameters of our model is multiple times of the other model. However, our model is still much faster than other models, which indicates that the assembled convolution does not occupy many computing resources, and most of the computing budget is still consumed by the conventional convolution on the feature maps.

Quantitative comparisons of our model and other efficient SR models are demonstrated in Tab.\ref{tab:2}. Our method achieves the best scores on every benchmark dataset. AsConvSR takes only 3.91 ms, which is close to the bicubic interpolation, and achieves an improvement on PSNR for more than 1dB. Our AsConvSR is the only method that achieves real-time performance with 1080p inputs. As demonstrated in Tab.\ref{tab:1}, AsConvSR-L exceeds RTSRN in PSNR with only 64.90\% of runtime.

\subsection{Visual Comparison}
\begin{figure}[t]
\centering
    \begin{subfigure}[b]{0.15\textwidth}
        \includegraphics[width=\textwidth]{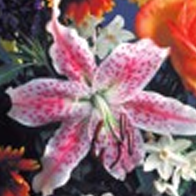}
        \includegraphics[width=\textwidth]{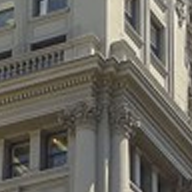}
        \caption{Bicubic}
    \end{subfigure}
    \begin{subfigure}[b]{0.15\textwidth}
        \includegraphics[width=\textwidth]{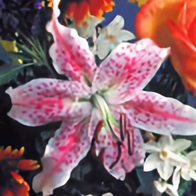}
        \includegraphics[width=\textwidth]{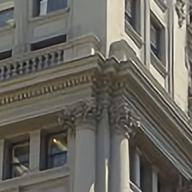}
        \caption{RTSRN}
    \end{subfigure}
    \begin{subfigure}[b]{0.15\textwidth}
        \includegraphics[width=\textwidth]{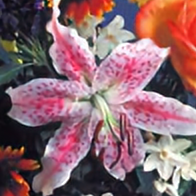}
        \includegraphics[width=\textwidth]{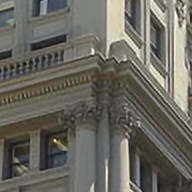}
        \caption{AsConvSR}
    \end{subfigure}
\caption{Visual results of AsConvSR and RTSRN on Set14\cite{zeyde2012single} and Urban100\cite{huang2015single}.}
\label{fig:7}
\end{figure}

Fig.\ref{fig:6} demonstrates the visual results of other efficient SR models and our AsConvSR-L. AsConvSR-L outperforms RTSRN in restoring compressed textures, and shows competitive performance in sharpness compared with RLFN\cite{kong2022residual} and FMEN\cite{du2022fast}. Fig.\ref{fig:7} shows the visual results of AsConvSR. The visual performance of AsConvSR significantly surpasses bicubic interpolation by using only 3.91 ms, and achieves similar quality to RTSRN. In general, both AsConvSR-L and AsConvSR keep the restoration accuracy and significantly reduce the model runtime.

\subsection{Ablation Study}
In this section, we evaluate the relationship between fidelity improvement and runtime consumption of various network structures. The ablation study is performed on the DF2K dataset, and Eq.\ref{eq:8} is used as the efficiency criterion.

\begin{table}[h]
\begin{center}
\begin{tabular}{cccc}
    \hline
    Method & Runtime & PSNR & Score \\
    \hline
    RLFN & 97.98 & 31.88 & 8.47 \\
    w/o ESA & 77.91 & 31.83 & 9.17 \\
    repeat upscaling & 75.78 & 31.82 & 9.21 \\
    \hline
\end{tabular}
\caption{Ablation study on the Enhanced Spatial Attention (ESA) module and repeat upscaling.}
\vspace{-0.5cm}
\label{tab:3}
\end{center}
\end{table}
We verify the efficiency of ESA and repeat upscaling using RLFN as the baseline. As demonstrated in Tab.\ref{tab:3}, adding ESA does improve network performance on PSNR, but it also increases the running time by 20.07 ms, which is 25.72\% of the runtime for the whole model. This phenomenon shows that ESA is not efficient enough with large resolutions inputs. Then we replace the bicubic interpolation in the model with repeat upscaling, which reduces the runtime by 2.13 ms. Given the standard of real-time 30 ms, using repeat upscaling can save 7.1\% of runtime, indicating that repeat upscaling is practical and efficient.

\begin{table}[h]
\begin{center}
\begin{tabular}{cccc}
    \hline
    Method & Runtime & PSNR & Score \\
    \hline
    Pixel-Unshuffle(1) & 22.58 & 31.39 & 12.53 \\
    Pixel-Unshuffle(2) & 15.72 & 31.33 & 14.46 \\
    Pixel-Unshuffle(3) & 15.49 & 31.23 & 13.30 \\
    Pixel-Unshuffle(4) & 12.56 & 31.05 & 13.31 \\
    \hline
\end{tabular}
\caption{Comparison of the pixel-unshuffle factor. Values in the brackets indicate different factors.}
\vspace{-0.5cm}
\label{tab:4}
\end{center}
\end{table}
The comparisons of the pixel-unshuffle factor are given in Tab.\ref{tab:4}. 'Pixel-Unshuffle(1)' means taking images of the original resolution as the input for the network, without using pixel-unshuffle to sample extra channel data from height and width. 'Pixel-Unshuffle(x)' means to sample pixels with interval of x from the height and width dimensions to the channel dimension. A larger x results in a larger number of channels. Therefore, for the experiments in Tab.\ref{tab:4}, the number of channels for the corresponding networks are 64, 128, 192, and 256 respectively, which can ensure the consistence of FLOPs in each experiment.

As demonstrated in the Tab.\ref{tab:4}, although the number of FLOPs is consistent in each experiment, the runtime and PSNR both decrease when the pixel-unshuffle factor increases. First, the performance in pixel-unshuffle experiment indicates that convolutions with a larger channel size and a smaller resolution run faster in NVIDIA GPU even the FLOPs remains unchanged. Second, pixel-unshuffle disrupts the spatial distribution of features and deteriorate the performance of the model. Finally, we choose 2 as the pixel-unshuffle factor in the following experiments.

\begin{table}[h]
\begin{center}
\begin{tabular}{cccc}
    \hline
    Method & Runtime & PSNR & Score \\
    \hline
    residual & 15.72 & 31.33 & 14.46 \\
    \hline
    w/o residual & 15.21 & 31.34 & 14.79 \\
    w/o global skip & 15.16 & 31.24 & 13.83 \\
    w/o residual\&bias & 14.14 & 31.35 & 15.43 \\
    \hline
\end{tabular}
\caption{Ablation study on the residual structure, bias and global skip connection.}
\vspace{-0.5cm}
\label{tab:5}
\end{center}
\end{table}

As demonstrated in Tab.\ref{tab:5}, we re-evaluate the efficiency of residual structure and bias of convolution. For the experiments without residual structure, we increase the values of certain kernel in convolution by 1 in the initialization phase. According to the theory proposed by RepOpt\cite{ding2022re}, such modification on initialization is equivalent to adding a skip connection to the corresponding convolution layer without explicitly adding this structure.

By removing the residual structure in the network, the performance is maintained and the runtime is decreased. On the contrary, global skip connection demonstrates its necessity as the PSNR drops greatly after the removal of global skip connection. In addition, training also becomes unstable after the removal of global skip connection. In conclusion, it is not preferable to remove the global skip connection from the super-resolution model. We also find that removing bias can improve the efficiency of the network. So we set up a new baseline without residual structure and bias for the following experiments.

\begin{table}[h]
\begin{center}
\begin{tabular}{cccc}
    \hline
    Method & Runtime & PSNR & Score \\
    \hline
    channels(128) & 14.14 & 31.35 & 15.43 \\
    channels(64) & 7.01 & 31.08 & 18.16 \\
    channels(32) & 3.73 & 30.69 & 19.04 \\
    channels(16) & 3.33 & 30.38 & 16.26 \\
    \hline
\end{tabular}
\caption{Comparison of models with different channel sizes.}
\vspace{-0.5cm}
\label{tab:6}
\end{center}
\end{table}
We compare the efficiency of the model with different channel sizes in Tab.\ref{tab:6}. The FLOPs and runtime do not proportionally decrease or increase when the number of channels changes. The FLOPs of a 128-channel network is almost four times that of a 64-channel network, but the runtime is only two times. However, considering the deterioration speed on PSNR, the 64-channel network still wins the competition in the efficiency score. In addition, the speed increase caused by reducing the number of channels has a limit. On the Tesla V100 platform, the reduction in runtime is no longer obvious, when the network is reduced to 16 channels. Given consideration on both the runtime and PSNR, we consider 32 as the optimal number of channels for our models.

\begin{table}[h]
\begin{center}
\begin{tabular}{cccc}
    \hline
    Method & Runtime & PSNR & Score \\
    \hline
    w/o dynamic & 3.73 & 30.69 & 19.04 \\
    \hline
    dynamic & 3.84 & 30.83 & 20.62 \\
    assembled & 3.91 & 30.87 & 21.05 \\
    \hline
\end{tabular}
\caption{Comparison between dynamic convolution and assembled convolution.}
\label{tab:7}
\end{center}
\end{table}
Comparisons on dynamic and assembled convolutions are shown in Tab.\ref{tab:6}. Replacing classical convolution with dynamic convolution only slightly increases the runtime (0.11 ms), but improves the PSNR by 0.14dB, which verifies the efficiency of the idea of divide-and-conquer. Compared with dynamic convolution, assembled convolution can keep the runtime basically unchanged while improving the final score by 0.43, achieving highest efficiency.

\begin{figure}[h]
\centering
\includegraphics[width=8.4cm, height=6.4cm, angle=0, scale=1]{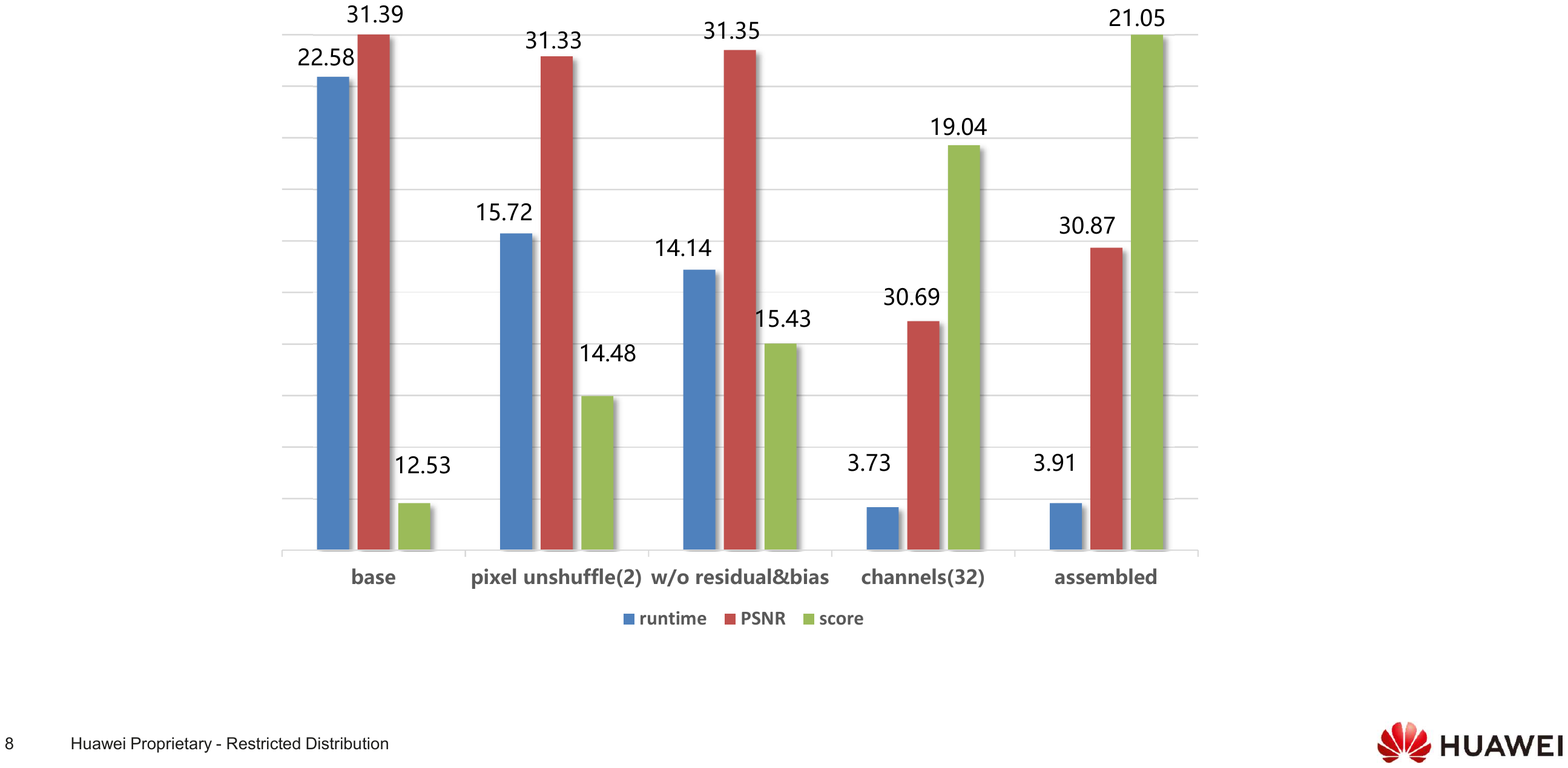}
\caption{Comparison of models with different efficient designs. 'base' is a vgg-style model with 64 channels and repeat upscaling. }
\label{fig:8}
\end{figure}
To compare the efficiency gains of each design, we collect all above experiments results into Fig. \ref{fig:8}. Adopting pixel-unshuffle, removal of residual and bias does not deteriorate the performance of the model in PSNR, and reduces runtime at the same time. Reducing the number of channels in the network increases the score by greatly reducing the runtime, but also reduces the PSNR. Assembled convolution can improve the PSNR without increasing the runtime, resulting in a significant score improvement and achieving highest efficiency among all the models.

\subsection{AsConvSR for NTIRE 2023 Real-Time Super-Resolution Challenge}
\begin{table}[h]
\begin{center}
\begin{tabular}{cccccc}
    \hline
    Method & Runtime & \makecell[c]{PSNR/SSIM\\/PSNR-Y} & Score \\
    \hline
    Bicubic & -- & 33.92/0.8829/36.66 & -- \\
    \hline
    Baseline\cite{conde2023ntire_rtsr} & 7.09 & 34.22/0.8854/-- & 9.27 \\
    DFCDN Team & 4.67 & 34.63/0.8916/37.46 & 15.17 \\
    Team OV & 2.91 & 34.62/0.8899/37.45 & 19.06 \\
    RTVSR & 2.24 & 34.71/0.8910/37.50 & 23.13 \\
    ALONG & 1.91 & 34.63/0.8906/37.38 & 23.81 \\
    \hline
    AsConvSR & 3.19 & 35.02/0.8957/37.74 & 24.13 \\
    \hline
\end{tabular}
\caption{Ranking of the NTIRE 2023 Real-Time Super-Resolution track 1\cite{conde2023ntire_rtsr}. Top 5 methods are included.}
\vspace{-0.5cm}
\label{tab:8}
\end{center}
\end{table}
Our model wins the first place in NTIRE 2023 Real-Time Super-Resolution - Track 1 ($\times$2). The major difference in training the competition model is the training datasets. In the competition we use DIV2K\cite{agustsson2017ntire}, Flick2K\cite{lim2017enhanced}, DIV8K\cite{gu2019div8k}, GTAV\cite{richter2016playing}, and LIU4K-V2\cite{Liu4K} for training. Other hyperparameters are consistent with the above experiments. As demonstrated in Tab.\ref{tab:8}, our AsConvSR exceeds the second best model in the PSNR score by 0.39 dB and is ranked first with the highest score.

\section{Conclusion}
In this paper, we propose a fast and lightweight super-resolution network with assembled convolution for real-time super-resolution. We revisit the efficiency of several designs such as pixel-unshuffle, repeat upscaling, residual and bias removal. Furthermore, we design a lightweight block named assembled block which can adaptively assembles the convolution kernels according to the input features. By introducing these designs, our model runtime is significantly reduced while an excellent super-resolution performance is obtained. Quantitative experiments demonstrate the competitive performance of our model.


{\small
\bibliographystyle{ieee_fullname}
\bibliography{egbib}
}

\end{document}